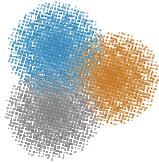

**Anura3D**
MPM Research Community

# Scalable and modular material point method for large-scale simulations

Krishna Kumar[1], Jeffrey Salmond[2], Shyamini Kularathna[3], Christopher Wilkes[1], Ezra Tjung[3], Giovanna Biscontin[1], Kenichi Soga[3]

[1] Department of Engineering, University of Cambridge, Cambridge, UK
[2] Research Software Engineering, University of Cambridge, Cambridge, UK
[3] Department of Civil Engineering, University of California, Berkeley, USA

[*] E-mail: kks32@cam.ac.uk

**ABSTRACT**

In this paper, we describe a new scalable and modular material point method (MPM) code developed for solving large-scale problems in continuum mechanics. The MPM is a hybrid Eulerian-Lagrangian approach, which uses both moving material points and computational nodes on a background mesh. Solving large-deformation problems such as landslides result in the material points actively moving through the mesh. Developing an efficient parallelisation scheme for the MPM code requires dynamic load-balancing techniques for both the material points and the background mesh. This paper describes the data structures and algorithms employed to improve the performance and portability of the CB-Geo MPM code.

**INTRODUCTION**

The Material Point Method (MPM) is a hybrid Eulerian-Lagrangian approach, which uses both moving material points and computational nodes on a background mesh. The MPM has been successfully applied to solve large-deformation problems such as landslides, failure of slopes, concrete flows, etc (Soga 2015). In the MPM, a continuum domain is discretized into a set of material points, which represent the material and carry all the updated information such as velocities, strain, stresses, and history variables. The main governing equations are solved at the nodes of a background computational mesh that covers the full problem domain and typically remains fixed throughout the calculation. Information stored at the nodes are typically reset at each time step. Variables required at the mesh to solve the governing equations are transferred from the material points to the nodes using mapping functions. The same mapping functions are used to update the quantities carried by the material points by interpolation of the mesh results.

As a hybrid Eulerian-Lagrangian method, the MPM is computationally expensive as it necessitates a frequent transfer of information between the nodes and the material points. Researchers have developed different parallelisation strategies varying from shared-memory parallel implementation (use of OpenMP in Anura3D, https://www.anura3d.com) to distributed multi-GPU parallelisation strategy (Dong and Grabe., 2018) to improve the computational performance of the MPM. By comparing the strong and weak scaling characteristics, Ruggirello and Schumacher (2014) observed the ghost nodes parallelisation strategy to be superior over ghost particles approach for scaling to large-scale problems.

Developing an efficient parallelisation scheme for a large-scale MPM code requires efficient vectorisation on modern architectures and dynamic load-balancing techniques for both the material points and the background mesh. This paper describes the data structures and algorithms employed to improve the performance and portability of the CB-Geo MPM code. The CB-Geo MPM code is developed as an open source project distributed under the MIT licensed and is available at https://github.com/cb-geo/mpm. The documentation and validation of the code can be viewed at https://cb-geo.github.io/mpm-doc/#/ and the benchmark problems are available at https://github.com/cb-geo/mpm-benchmarks.

# CODE OVERVIEW

The objectives of developing the CB-Geo MPM code are the following: (a) to model large-scale multiphase multi-physics problems in a reasonable time, (b) develop a modular system that can handle different MPM algorithms (explicit/implicit/projection method), material models, and mapping functions (linear/GIMP/CPDI) and (c) the ability to model complex geometries and boundary conditions and (d) efficient, easy-to program parallelisation and scaling strategy for large-scale simulations. The CB-Geo MPM code is a generic template-based modern C++14 code base. An object-oriented programming paradigm is adopted to modularise the MPM code. The Unified Modelling Language (UML) diagram of the CB-Geo MPM code is shown in Figure 1.

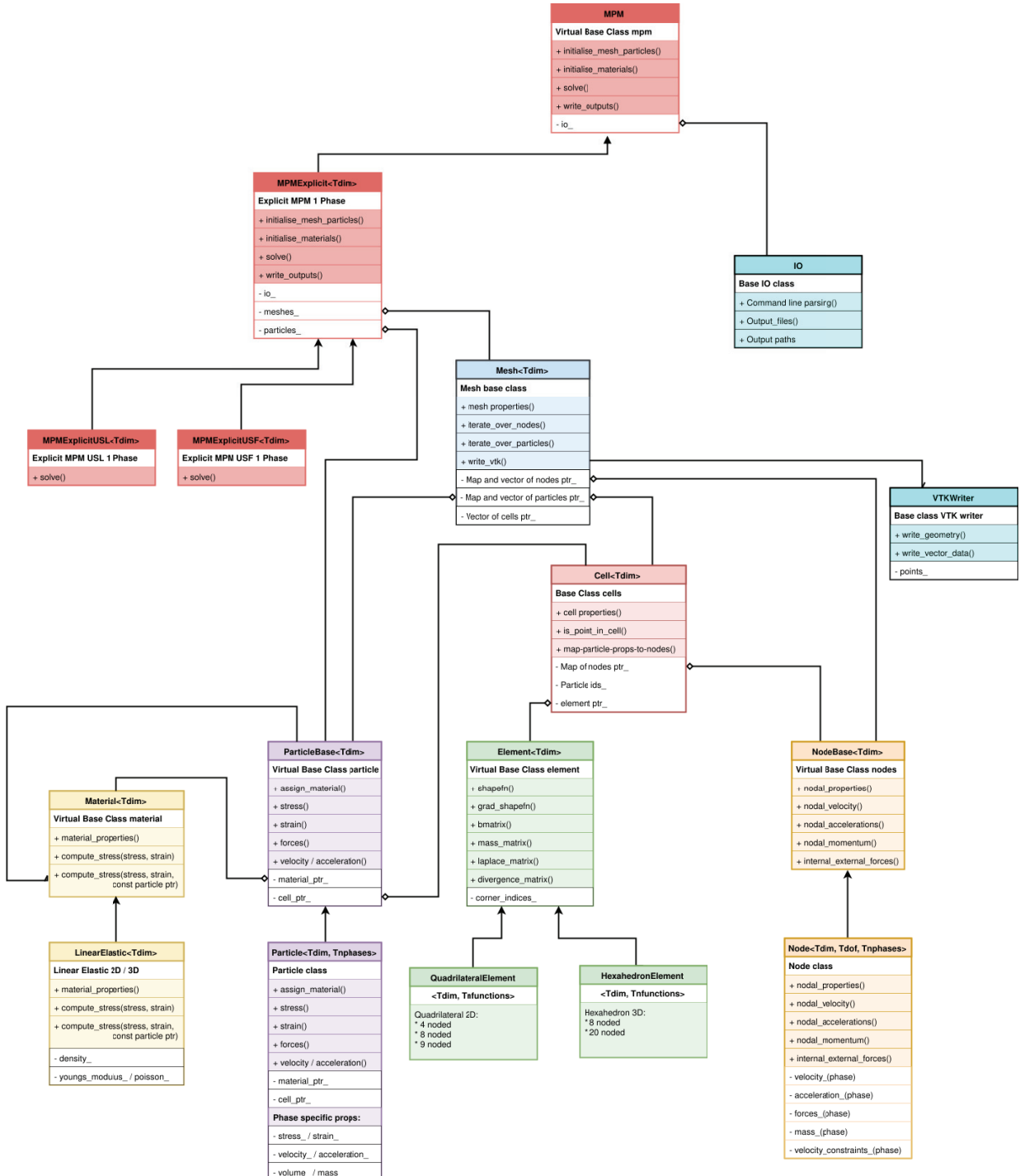

Figure 1 A simplified UML diagram of the CB-Geo MPM code structure

In order to develop a modular strategy for different entities in the MPM code such as element types, material models, particle types (soil/water), etc. the factory creation pattern is adopted. A generic factory pattern is used to handle constructors with an arbitrary number of arguments to instantiate material points, nodes and cells of required types. This requires defining an abstract class and concrete classes for each entity. For example, an abstract base class of element is used to create concrete element types of quadrilateral and hexahedral elements.

The background grid is defined using the `Node` and `Cell` class objects. The `Node` class stores properties such as velocity, acceleration, forces, mass, and momentum for each phase. These properties are reset at each time-step, unlike frictional and velocity constraints. Each `Cell` class owns a vector of shared-pointers to the `Node` objects and a shared-pointer to its type (`Element` class object). The `Element` class defines the type of the element (4, 8 and 9-noded Quadrilateral or 8, 20 and 27-noded Hexahedron) and the associated shape functions (Linear/GIMP/CPDI). A `Cell` and its associated `Node` objects are considered active during a simulation if the `Cell` has at least one active material point. The material points (`Particle` class objects) are treated as the fundamental unit in the MPM code and has a owning shared-pointer to the `Cell` class, which is initialised when a material point can be located in the cell. The information transfer between the material points (`Particle`) and the corresponding `Node` objects are handled via the `Cell` associated with each material point. All material points have an associated shared-pointer to an abstract `Material` class object, which is typically used to calculate the updated stresses for a given strain tensor. An overarching `Mesh` class has a container of an associated sub-domain of the mesh objects (`Node` and `Cell`) and the material points at that given time-step. A factory method pattern is also used to abstract different MPM algorithms. The CB-Geo MPM code currently supports Update Stress First (USF) and Update Stress Last (USL) explicit algorithms.

**DATA STRUCTURES AND PARALLELISATION STRATEGY**

Data structures offer a particular way of organising and storing information in a computer so that it can be accessed and modified efficiently. For efficient large-scale optimisation of the MPM, it is critical to choose the appropriate data structure for material properties such as stresses, velocity, acceleration, etc and at the `Mesh` level to store the list of `Particle`, `Cell` and `Node` objects associated with the sub-domain.

An Eigen Matrix (array) structure is used to store tensor quantities such as location, velocities, acceleration, stresses, strains, etc. The size of these tensor quantities depend on the dimension of the problem (for e.g., coordinates), or it is fixed (for e.g., the stress tensor has 6 components in the Voigt notation). Typically, updating these tensor quantities require a `for` loop, and a loop over a known array size can be unrolled and optimised during compilation. However, the flexibility of modelling different phases and dimensions using a single material point. For example, using a 2D or 3D array based on the problem dimensions, and a multiphase continuum discretized using a single point multiphase formulation requires a dynamically sized array, with each column in a matrix representing a different phase. Dynamically sized arrays cannot be unrolled at compile-time and are often less efficient. To improve the compile-time optimisation, the `Particle` class uses templates to define the dimension and number of phases, and the `Node` class uses an additional template argument of degrees of freedom. Templates in C++ offers abstraction and also enables optimisation as the sizes of the arrays are known at compile time. Thus, the vectorisation capability of modern architectures is exploited by having a fixed size array during compilations.

Modern CPUs are designed for Single Instruction Multiple Data (SIMD) operations and vectorizations. The `Mesh` class uses a Intel Threaded Building Blocks parallel vector container to store and operate on the material points (`Particle`), `Cell` and `Node` objects in the sub-domain. The Intel TBB is a C++ template library for task-based parallel execution model on multi-core processors. The TBB `concurrent_vector` container spawns and schedules threads to execute the tasks on each element of the container in parallel. The material point properties such as acceleration, velocities, and positions can be updated independently of other material points in the mesh and hence can be executed in parallel without the need for any special handling. On the other hand, nodes share information with all the material points in the associated cells (and also the neighbouring cells in the case of GIMP). Updating this nodal information (for e.g., velocity/momentum from the material point) requires aggregating properties from all the material points in the associated cell and its neighbours. The nodal information cannot be updated independently, as a node may be required to be updated by multiple material points at a given time-step. When the nodal information is updated in parallel without any special barriers, this may result in a thread updating the information while another thread (material point object) is reading a previous value of the nodal tensor quantity - this phenomenon is called a race condition. In order to effectively use the parallel containers and spawn tasks in parallel, a special locking mechanism is necessary. To prevent race conditions on nodal updates, an associated mutex lock is enabled at each `Node` class during a property update. Thus, all functions on material points and nodes in the code are parallel operations, with atomic updates for `Node` class functions. This exploits the shared-memory parallel architecture. Additionally, `task_groups` are used to spawn independent tasks that can be run in parallel, for example, computing the external and internal forces at the nodes can be independent tasks running in parallel. A task group barrier `wait()` is used to synchronise these calls.

Although parallel vector containers are useful to vectorise, fetching a particular node or a material point requires iterating through all the components in a container and has the worst performance order of O($n$), where '$n$' is the number of elements in the container (size). To improve the performance of identifying a material point pointer or a nodal pointer, a hashed map data structure is utilised. Although a traditional hashed-map has an average order of lookup O(1), it might suffer for hash collisions, that degrades the performance of the container even for a small number of elements. The Robin Hood hashing algorithm minimises the hash collisions and has a worst-case performance of O($ln\ n$). A Robin Hood hashed map is created with the unique global id of the node/material points as the key and a shared-pointer as its value. The map container is used to identify and apply functions (for e.g., boundary conditions) at a given node or a material point.

In order to solve real-scale landslides, it is not sufficient to merely utilise the shared-memory architecture, but it is important to decompose the domain across several compute nodes to take advantage of high-performance computing clusters. The CB-Geo MPM exploits the multiple-CPU parallel framework with Message Passing Interface (MPI). Typically, in the MPM, the number of particles in a problem is at least an order of magnitude higher than the number of cells representing the background grid (for example 16 material points per cell is typically used in 2D simulations). To improve the load-balance across compute nodes, the material points in a mesh are split across compute nodes equally, and any residual material points are allocated to the MPI rank 0. Since the memory requirement for the background grid is significantly smaller than the memory requirement of the material points (a typical ratio of number of points to the cell is 16 in 2D and 64 in 3D for 4 material points in each direction), the background grid is copied across different MPI tasks. Two parallel `all_reduce` functions are used to update the scalar and tensor nodal properties. A hybrid MPI+X parallel strategy is used to update nodal information across compute nodes. Although in this scheme all the nodes in the mesh need to be updated across compute nodes, the `all_reduce` operation across the MPI ranks are optimised in addition using shared-memory parallel tasks. The code can be further improved by using sub-domains and updating the nodal information only on the nodes (and halo nodes for GIMP/CPDI neighbouring cells) shared between subdomains (ghost nodes).

**ISOPARAMETRIC MESH**

In the MPM, the velocity constraints and acceleration (frictional) boundary conditions are applied at the nodes. Non-prismatic boundaries like those that can be found in landslides pose problems when applying nodal boundary conditions on irregular surfaces. In addition to a Cartesian grid, the CB-Geo MPM code uses isoparametric elements to model complex geometries and boundary conditions. One of the critical performance bottleneck of the MPM is the location of material points in a mesh, as the material points traverse freely through the grid as the body deforms. Purin (2016) observed that using a neighbourhood algorithm based on the magnitude of the displacement reduces the compute time of locating the particles on a Cartesian grid by around 75 to 95% in comparison to a brute force search.

Although the use of isoparametric elements in the MPM allows for modelling non-prismatic geometries more accurately, it poses a unique problem. Unlike the Finite Element Method, where the location of Gauss points in an unit cell is known, the use of isoparametric elements in the MPM requires transforming the location of material points from the real coordinates to the natural coordinates. The inverse transform of the linear mapping from natural to real coordinates does not have an analytical solution in 3D.
An efficient approach to transform a material point from the real cell to a unit cell is using an iterative approach like Newton Raphson. This inverse transformation operation is computationally very intensive. The efficiency of an iterative approach depends on the initial guess. In 2D, An inverse bilinear interpolation technique is used in conjunction with Newton Raphson for transforming the point. Additionally, the Affine transformation is used when linear mapping functions are used to make an improved initial guess. This combined approach in transforming real to natural coordinates reduces the computational cost. At each time-step, the material point is first checked for its location in the current cell. This solution can be further improved by adopting a list of neighbours, in subsequent steps of the simulation when the magnitude of deformation of each material point is known. The function to locate particles in a mesh uses thread based parallelisation across particles. The performance of the function to locate 150,000 particles in a mesh of 23,000 cells improved from 14 seconds on a single thread to 2.7 seconds for 32 threads, with a speed-up of 5x on 32 shared-memory cores. The performance of this function will increase with an increase in the number of particles and cells in a mesh.

**BENCHMARK AND PERFORMANCE**

The stress concentration around a plate with a circular hole subjected to uniform pressure is used as a validation example for the CB-Geo MPM code. A plane strip (plane stress) of width 5 m, length 8 m with a centred hole of radius 0.5m is subjected to uniform tension of $\sigma_0 = 100$ Pa applied at the ends of the strip. Due to symmetry, a quarter of this plate is modelled. For a finite width plate, the elastic stress state depends on the plate width and the hole radius. Howland (1930) proposed stress-ratios for a range of geometries based on a semi-analytical solution of the elastic stresses at the circular hole. The MPM simulations are carried out using the Explicit USF algorithm. The background grid comprises of cells of size 0.0625m, and the plate is discretized with 40,156 material points, with 16 material points per cell. The stress distribution from the MPM simulation is shown in Figure 2. It can be seen from Table 1 that the stress concentration factors at the edge of the plate hole matches the analytical solution.

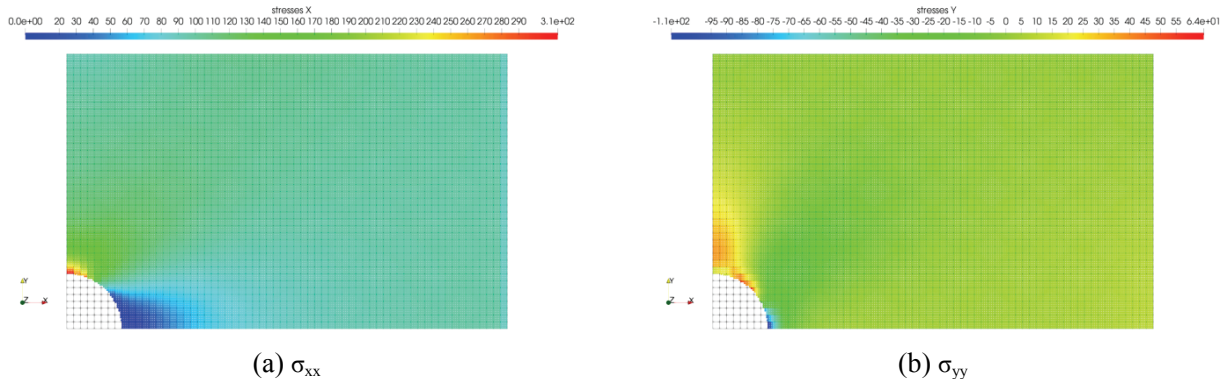

(a) $\sigma_{xx}$          (b) $\sigma_{yy}$

Figure 2 MPM simulation of stresses around a circular hole on a plate subjected to axial loading

Table 1 Comparison of stress concentration factors between the MPM simulations and the semi-analytical solution of a plate with a circular hole subjected to uniform tension.

| Stress concentration factors | Howland's Semi-analytical solution | CB-Geo HPC MPM (USF) |
|---|---|---|
| Maximum $\sigma_{xx}/\sigma_0$ | 3.14 | 3.13 |
| Minimum $\sigma_{xx}/\sigma_0$ | 0.00 | -0.029 |
| Maximum $\sigma_{yy}/\sigma_0$ | -1.11 | -1.08 |

The strong-scaling characteristics of the benchmark problem of a plate with a circular hole are evaluated on a shared-memory node to understand the performance of the CB-Geo MPM code. The scaling tests were run using a Singularity container (https://cloud.sylabs.io/library/cbgeo/mpm) on a 2x Intel Xeon Skylake 6142 processors with 2.6GHz 16-cores at the CSD3 HPC cluster at Cambridge. The compute time for 1000 steps using MPM Explicit USF algorithm decreases from 59.28 +/- 2.30s for a single-thread execution to 18.08 +/- 0.44s for 16 threads (see Figure 3). The performance does not improve beyond 16 threads, due to the overheads involved in spawning threads for a small number of material points. Further scaling tests are required to evaluate the strong and weak scaling characteristics of the MPM code on both distributed and shared-memory architectures to evaluate the capability of the MPM code for simulating real-scale landslides.

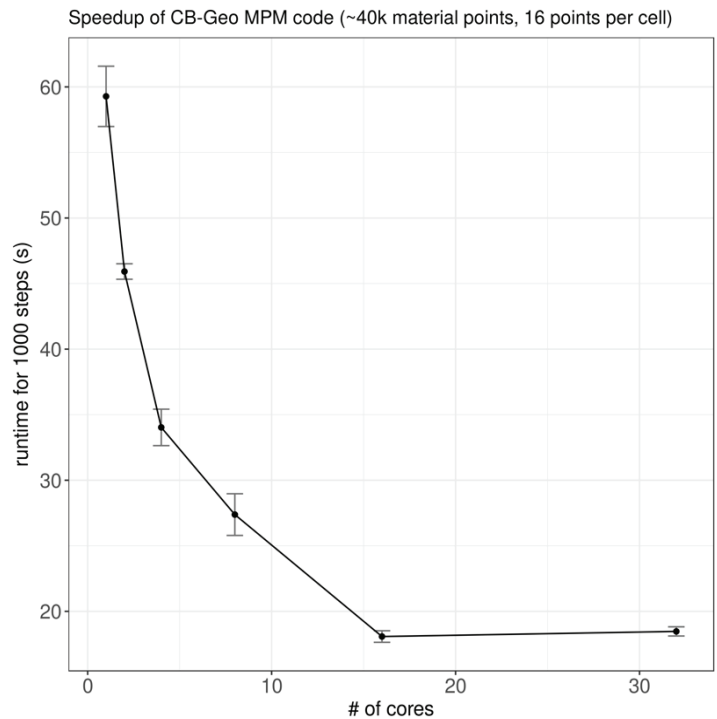

Figure 3 Runtime for the MPM simulations of stresses around a circular hole on a plate subjected to axial loading with 40,156 material points in a mesh with 16 material points per cell. The simulations were run on a Singularity container on a 2x Intel Xeon Skylake 6142 processors with 2.6GHz 16-cores

**SUMMARY**
An open source, scalable, modular, and an extensible 2D/3D explicit MPM code is developed in the present study capable of solving field-scale simulations. One of the challenges when solving large deformation problems using the MPM is that the material points are actively moving through the computational grid. We adopt a combined MPI+X (Intel TBB) multi-CPU parallelisation strategy with load-balancing techniques for efficient real-scale simulations. The performance of the code is validated by analysing an elastic plane stress problem. The results of the strong-scaling tests demonstrate the scalability of the code for shared-memory architecture. Further, an isoparametric mesh is implemented in the code to handle complex geometries such as a topology of a real-scale landslide. The issue of transforming material points from a real cell to a unit cell with natural coordinates is achieved using a hybrid Newton Raphson iterative approach with Affine transformation and an inverse bilinear interpolation. Numerical results for complex large-scale problems will be presented elsewhere. Further research will be carried out to showcase the performance characteristics of the CB-Geo MPM code for real-scale problems.